\documentclass[pra,twocolumn]{revtex4-2}
\usepackage{eurosym}
\usepackage{amssymb}
\usepackage{amsfonts}
\usepackage{amsmath}
\usepackage{graphicx}
\usepackage{bm}
\usepackage{braket}
\usepackage{mathtools}
\usepackage{textcomp}
\usepackage{xcolor}
\usepackage{float}

\begin{document}

\title{Square-shaping of sturdy optical vortex droplets in
	quasi-phase-matched photonic crystals}
\author{Xuening Wang$^{1}$}
\author{Yimin Zhang$^{1}$}
\author{Qiuyi Ning$^{1,2}$}
\author{Bin Liu$^{1,2}$}
\author{Li Zhang$^{1,2}$}
\author{Hexiang He$^{1,2}$}
\email{sysuhhx@163.com}
\author{Boris A. Malomed$^{3,4}$}
\author{Yongyao Li$^{1,2}$}
\email{yongyaoli@gmail.com}
\affiliation{$^1$School of Physics and Optoelectronic Engineering, Foshan University,
Foshan 528225, China\\
$^2$Guangdong-Hong Kong-Macao Joint Laboratory for Intelligent Micro-Nano
Optoelectronic Technology, Foshan University, Foshan 528225, China\\
$^3$Department of Physical Electronics, School of Electrical Engineering,
Faculty of Engineering, Tel Aviv University, Tel Aviv 69978, Israel\\
$^4$Instituto de Alta Investigaci\'{o}n, Universidad de Tarapac\'{a},
Casilla 7D, Arica, Chile}

\begin{abstract}
We elaborate a scheme for controllable shaping of self-trapped vortex states
in a quasi-phase-matched three-dimensional photonic crystal with the
combination of self-focusing quadratic and defocusing cubic material
nonlinearities. The setting gives rise to sturdy droplet-like vortex modes,
capable to adapt to externally imposed strong geometric constraints. The
application of a square-shaped modulation in the transverse $\left(
x,y\right) $ plane and periodic quasi-phase-matching to the quadratic
nonlinear coefficient $d_{z}$ along the propagation direction $z$ leads to
the formation of square vortex droplets (VDs) with fourfold rotational
symmetry ($C_{4}$). These states preserve the vortical phase circulation and
exhibit robust propagation in a broad parameter region. In the oversaturated
regime dominated by the cubic self-defocusing, the square-shaped VDs obey
the anti-Vakhitov--Kolokolov stability criterion. The results, which are
produced, chiefly, for the VDs with topological charge $S=1$, and also, in a
partial form, for $S=2$ and $4$, reveal an unexpected universality:
the vortex robustness is not contingent upon the circular symmetry. Thus, the
combination of the competing nonlinearities and geometric confinement provides not only
an effective method for the formation of self-trapped vortex states, but also new
insight into generality of the topological protection in nonlinear optical fields.
\end{abstract}

\maketitle


\section{Introduction}

Vortex states constitute a class of topological excitations in nonlinear
wave systems \cite{gprl69}, distinguished by the carried orbital angular
momentum \cite{lpra45,mpt57,slpr2,jnp6} and intrinsic phase circulation \cite%
{jprsa336,ylsa8,apu65}. In conventional nonlinear systems, dominated by
quadratic or cubic ($\chi ^{(2)}$ or $\chi ^{(3)}$) self-focusing
nonlinearities, bright solitons with embedded vorticity, alias vortex rings,
are subject to splitting instability, which breaks the vortex ring into a
set of fragments \cite{Desyatnikov,baip2022}. In $\chi ^{(3)}$ nonlinear
media, the splitting instability is related to the critical or supercritical
collapse of bright solitons in the 2D and 3D geometry, respectively \cite%
{gsc2015,Torner}. In media governed by the $\chi ^{(2)}$ nonlinearity, the
collapse does not occur, hence fundamental (zero-vorticity) 2D and 3D
solitons are stable; nevertheless, their vortex counterparts are unstable
against the spontaneous splitting initiated by the azimuthal modulation
instability of the vortex rings. As a result, they suffer the fragmentation
into a set of separating fundamental solitons \cite%
{apr2002,wprl1995,wol1995,wprl1997,lel1997,dol1998}. Systems with competing
nonlinearities offer a possibility to stabilize vortex solitons against the
splitting \cite%
{ipre2001,dpre2004,pprl2000,dcsf2025,rol1992,cprl1995,soc1996,aol1995,dpre2000,pego,paz,liyy,paredes,z208}%
. In particular, the stabilization of the vortex rings may be provided by
the competition of quadratic and self-defocusing cubic nonlinearities. In
this case, the wave field can self-trap into \textquotedblleft
liquid-like\textquotedblright\ states, often referred to as
\textquotedblleft droplets" \cite{kprl121,npra98,xoe2023,xpra556}. They
feature macroscopic properties reminiscent of classical liquids: while
maintaining a well-defined internal phase and vortex topological structure,
the overall shape is highly adaptive to an external potential or geometric
constraints.

The \textquotedblleft ductility" of the vortex-droplet (VD) states suggests
a question if it is possible to cast their transverse shape into a desirable
form, while preserving the intrinsic vorticity and propagation stability.
This issue has been partially explored in matter-wave systems, such as
Bose--Einstein condensates \cite%
{dprl115,dprl117,mn539,cs359,ybpra,yzpra,zfop16,zns93,yfop16,gfop19}. It has
been shown, in particular, that, in multicomponent systems or those
dominated by long-range dipole--dipole interactions, VDs can evolve into
anisotropic or even noncircular stationary configurations which keep the
intrinsic topological charge (i.e., the embedded vorticity) \cite%
{gprl133,gz112}. Those findings suggest that self-bound optical droplets may
be able to maintain stable propagation while undergoing substantial shape
deformation, thereby providing a versatile platform for the control of field
morphologies.

Motivated by this background, we propose and develop a novel scheme for VD
shaping, based on the use of quasi-phase-matched (QPM) photonic crystals. By
introducing the competing quadratic $\chi ^{(2)}$ and cubic $\chi ^{(3)}$
nonlinearities, the system admits the formation of stable VD states. The QPM
photonic crystal is engineered to incorporate a square-shaped photonic
potential in the transverse $(x,y)$ plane as a pattern-forming element,
while the local $\chi ^{(2)}$ coefficient is periodically modulated along
the propagation direction to achieve the QPM. This design deliberately
breaks the continuous rotational symmetry of the system, creating favorable
conditions for the stabilization of VD modes with discrete (rather than
continuous rotational) symmetry. The choice of the C4 symmetry is not arbitrary:
it is the fundamental rotational symmetry of the underlying square-lattice photonic crystal,
which is directly inspired by our previous work on discrete square-shaped vortex solitons in
purely quadratic media \cite{xypra}. Thus, C4 serves both as an experimentally relevant
testbed and as a natural starting point for exploring the discrete-symmetry shaping. In this framework, we demonstrate that an
initially circular VD soliton with a ring-shaped intensity profile readily
reshapes into a \textquotedblleft square VD\textquotedblright , featuring a
clear fourfold rotational symmetry, while maintaining the topological charge
and persistent propagation. We stress that this result reveals a
counter-intuitive yet universal feature of the vortex modes: their angular momentum and topological robustness are not contingent upon
the continuous rotational symmetry, persisting under the action of the broken rotational symmetry, here reduced to C4.

By synergistically combining the competing nonlinearities with the
engineered geometric constraints, this work puts forward an experimentally
feasible scenario for controlling the morphology of self-trapped optical
vortex modes. Beyond the morphology control, our findings provide the new
insight into the generality of the topological protection in nonlinear optical fields, showing that
the vortex droplets retain their essential assets, \textit{viz}., the topological charge and stable propagation, even when reshaped into a low-symmetry square geometry. The universality presents the fundamental interest and practical relevance, as it decouples vortex functionality from the strict circular symmetry and enables integration into platforms with anisotropic architectures. The proposed approach avoids the need for complex anisotropic
interaction mechanisms, being instead built upon the mature and precisely
fabricable platform of nonlinear photonic crystals \cite%
{fx130,xypra,yx112,tnp2018,dnp2018,snp2018,alsa2021,alpr2010,hfo2020,saom2023}%
. Thus, our results suggest a practical strategy for engineering nonlinear
optical fields with tailored spatial geometries and topological properties.

We introduce the theoretical model in Section 2, followed by presenting
results of the study of the model in Section 3. An estimate of relevant
experimental parameters is provided in Section 4. The paper is concluded by
Section 5.

\section{The model}

The paraxial propagation of light beams through the 3D QPM photonic crystals
with the competing $\chi ^{(2)}$ and $\chi ^{(3)}$ nonlinearities is
governed by the coupled equations for the slowly varying
fundamental-frequency (FF) and second-harmonic (SH) amplitudes, $A_{1}$ and $%
A_{2}$:
\begin{equation}
\begin{split}
i\partial _{Z}A_{1}=& -\frac{1}{2k_{1}}\nabla _{XY}^{2}A_{1}-\frac{%
2d(X,Y,Z)\omega _{1}}{cn_{1}}A_{1}^{\ast }A_{2}e^{-i\Delta k_{0}Z} \\
& +\frac{3\chi ^{(3)}\omega _{1}}{2cn_{1}}(|A_{1}|^{2}+2|A_{2}|^{2})A_{1}
\end{split}
\label{e1}
\end{equation}

\begin{equation}
\begin{split}
i\partial _{Z}A_{2}=& -\frac{1}{2k_{2}}\nabla _{XY}^{2}A_{2}-\frac{%
2d(X,Y,Z)\omega _{2}}{cn_{2}}A_{1}^{2}e^{i\Delta k_{0}Z} \\
& +\frac{3\chi ^{(3)}\omega _{2}}{2cn_{2}}(|A_{2}|^{2}+2|A_{2}|^{2})A_{2}
\end{split}
\label{e2}
\end{equation}%
where $\nabla _{XY}^{2}\equiv \partial _{X}^{2}+\partial _{Y}^{2}$ is the
paraxial-diffraction operator, $c$ is the speed of light in vacuum, while $%
n_{1,2}$, $\omega _{1,2}$ $(\omega _{2}=2\omega _{1})$, and $k_{1,2}$ are,
respectively, the refractive indices, carrier frequencies, and wavenumbers
of the FF and SH components, and $\Delta k_{0}=2k_{1}-k_{2}$ is the
phase-velocity mismatch. $\chi ^{(3)}>0$ is the third-order susceptibility,
which accounts for the cubic self-defocusing. The local modulation of the
second-order susceptibility $\chi ^{(2)}$ is determined by coefficient $%
d(X,Y,Z)=\sigma (X,Y)d(Z),$ with a square structure in the $(X,Y)$ plane, which naturally exhibits $C_{4}$ rotational symmetry,

\begin{figure}[tbp]
{\includegraphics[trim=105 50 130 80,clip,width=3.4in]{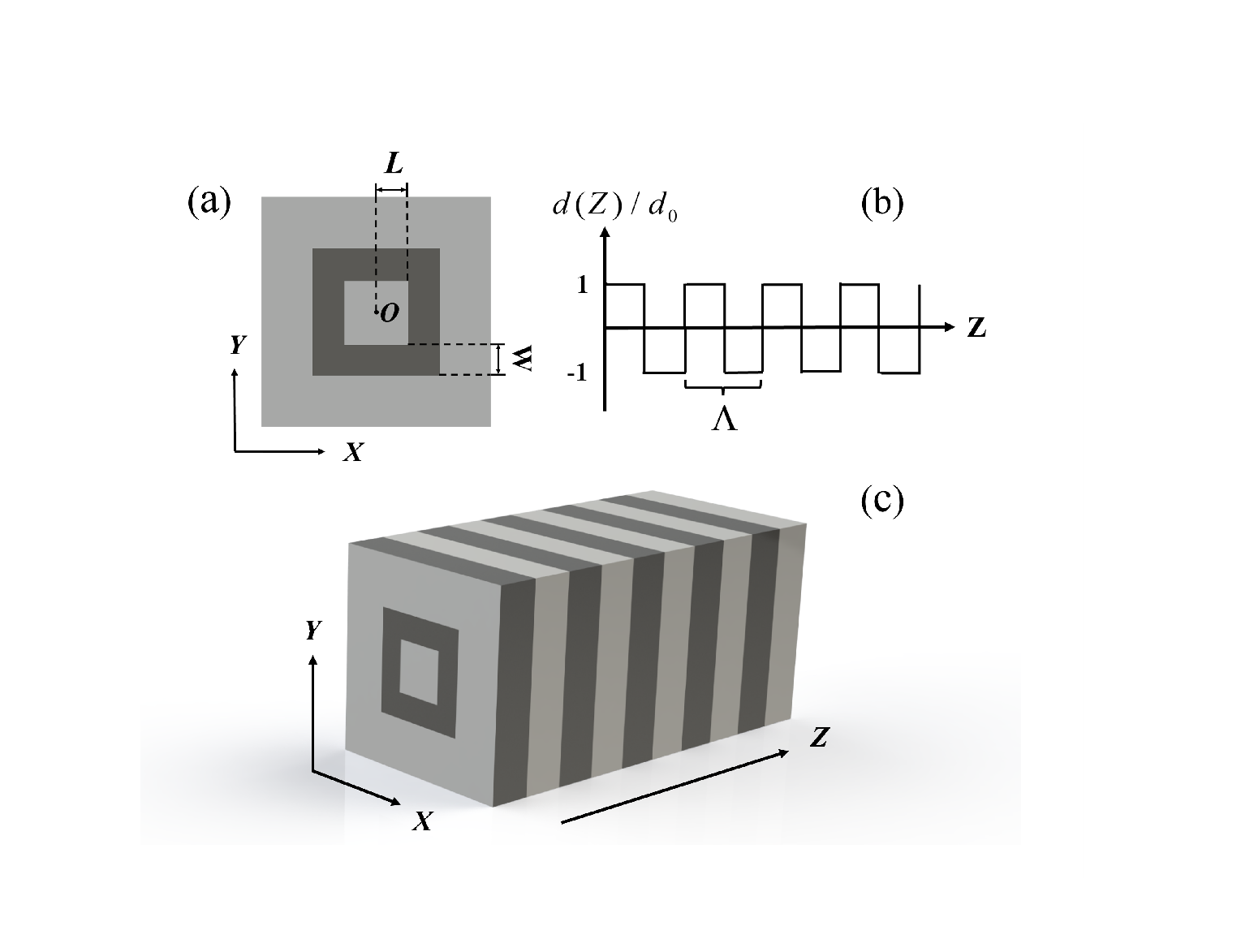}}
\caption{(a) A schematic illustration of the modulation profile of the
nonlinear photonic crystal in the $\left( x,y\right) $ plane. (b) The
periodic modulation along the $z$ axis with period $\Lambda $, as defined by
Eq.~(\protect\ref{d1}). (c) The 3D schematic of the quasi-phase-matched
photonic crystal.}
\label{f1}
\end{figure}

\begin{align}
\sigma _{x}(X,Y)& =\mathrm{sgn}\Big[\mathrm{sgn}\!\left( L+W-|X|\right)
\mathrm{sgn}\!\left( L+W-|Y|\right)  \notag \\
& \quad -\mathrm{sgn}\!\left( L-|X|\right) \mathrm{sgn}\!\left( L-|Y|\right) %
\Big].  \label{sigma}
\end{align}

According to Eq.~(\ref{sigma}), the square modulation frame is characterized
by side length $L$ and thickness $W$, as shown in Fig.~\ref{f1}(a). The
modulation function of the $\chi ^{(2)}$ coefficient along the propagation
direction is defined by a periodic structure,
\begin{equation}
d(Z)=d_{0}\mathrm{sgn}[\cos (2\pi Z/\Lambda )],  \label{d1}
\end{equation}%
with modulation period $\Lambda $ and amplitude $d_{0}$. The expression in
Eq.~(\ref{d1}) can be further represented by its Fourier expansion: \cite%
{aoe2018,afp2021}:
\begin{equation}
d(Z)=d_{0}\sum_{m\neq 0}\frac{2}{m\pi }\sin \left( \frac{m\pi }{2}\right)
\exp \left( i\frac{2{\pi }m}{\Lambda }Z\right) .  \label{d2}
\end{equation}%
see Fig.~\ref{f1}(b). As usual \cite{Torner,apr2002}, only the terms with $%
m=\pm 1$ are retained in Eq.~(\ref{d2}) in the subsequent analysis, as these
terms play the dominant role in the QPM effect.

By means of the transformation (cf. Refs. \cite%
{cjosab2013,foe2021,ypra2020,gjosab2000,jpla2021}),

\begin{equation}
\begin{split}
u_{j}=\frac{\chi^{(3)}}{d_0} \sqrt{\frac{n_j}{\omega_j \alpha}}\, A_j \exp\!%
\left[i\left(\Delta k_0 - \frac{2\pi}{\Lambda}\right) Z \right],\quad j=1,2,
\end{split}
\label{uj}
\end{equation}

\begin{equation}
\begin{split}
\alpha=(\frac{n_1}{\omega_1}+\frac{n_2}{\omega_2}), \quad{z_d}^{-1}=\frac{2{%
d_0}^2}{c{\pi}{\chi}^{(3)}}\sqrt{{\frac{{\omega_{1}^{2}}{\omega_2}{\alpha}}{{%
n_{1}^{2}n_2}}}},
\end{split}
\label{zd}
\end{equation}

\begin{equation}
z = \frac{Z}{z_d}, \qquad x = \sqrt{\frac{k_1}{z_d}}\,X, \qquad y = \sqrt{%
\frac{k_1}{z_d}}\,Y,  \label{xyz}
\end{equation}

\begin{equation}
\Omega = (\Delta k_0 - \frac{2\pi}{\Lambda}) z_d,  \label{omega}
\end{equation}

\begin{equation}
\gamma_{11} = \frac{3\pi}{4} \sqrt{\frac{\omega_1^{2} n_2 {\alpha}}{%
n_1^{2}\omega_2}},\qquad \gamma_{12} = \frac{3\pi}{2} \sqrt{\frac{\omega_2 {%
\alpha}}{n_2}},  \label{gamma1}
\end{equation}

\begin{equation}
\gamma_{22} = \frac{3\pi}{4} \sqrt{\frac{\omega_2^{3} n_1^{2} {\alpha}}{%
n_2^{3}\omega_1^{2}}},  \label{gamma2}
\end{equation}

Eqs. (\ref{e1}) and (\ref{e2}), in which, as said above, only the terms with
$m=\pm 1$ are kept in Eq.~(\ref{d2}), are cast in the simplified form:
\begin{equation}
\begin{split}
i\partial _{z}u_{1}& =-\frac{1}{2}\nabla ^{2}u_{1}-\Omega u_{1}-2\sigma (x,y)%
{u_{1}^{\ast }}u_{2} \\
& +({\gamma }_{11}|u_{1}|^{2}+{\gamma }_{12}|u_{2}|^{2})u_{1},
\end{split}
\label{eq1}
\end{equation}%
\begin{equation}
\begin{split}
i\partial _{z}u_{2}& =-\frac{1}{2{\eta }}{\nabla }^{2}u_{2}-\Omega
u_{2}-\sigma (x,y){u_{1}^{2}} \\
& +({\gamma }_{22}|u_{2}|^{2}+{\gamma }_{12}|u_{1}|^{2})u_{2},
\end{split}
\label{eq2}
\end{equation}%
where $\nabla ^{2}=\partial _{xx}+\partial _{yy}$ and $\eta =k_{2}/k_{1}$.
Neglecting the slight difference between the refractive indices of the FF
and SH components, i.e., setting $n_{1}=n_{2}$, the coefficients in Eqs.~(%
\ref{eq1}) and (\ref{eq2}) reduce to
\begin{equation}
\gamma _{11}=3\pi \sqrt{3}/8,\gamma _{22}=\gamma _{12}=4\gamma _{11},\eta =2.
\label{coeff}
\end{equation}

According to the Manley-Rowe relations \cite{gjosab2013}, the system
conserves two dynamical invariants, \textit{viz}., the total power,
\begin{equation}
P=\int \int \left( |u_{1}|^{2}+2|u_{2}|^{2}\right) dxdy  \label{power}
\end{equation}%
and Hamiltonian, composed of four terms:
\begin{equation}
H=\int \int \left( \mathcal{H}_{P}+\mathcal{H}_{\Omega }+\mathcal{H}_{2}+%
\mathcal{H}_{3}\right) \,dx\,dy,  \label{h}
\end{equation}
\begin{equation}
\mathcal{H}_{P}=\frac{1}{2}|\nabla u_{1}|^{2}+\frac{1}{4}|\nabla u_{2}|^{2},
\label{h1}
\end{equation}%
\begin{equation}
\mathcal{H}_{\Omega }=-\Omega (|u_{1}|^{2}+|u_{2}|^{2}),  \label{h2}
\end{equation}%
\begin{equation}
\mathcal{H}_{2}=\sigma (x,y)\left( {u_{1}^{\ast }}^{2}u_{2}+\mathrm{c.c.}%
\right) ,  \label{h3}
\end{equation}%
\begin{equation}
\mathcal{H}_{3}=\frac{1}{2}{\gamma }_{11}|u_{1}|^{4}+{\gamma }%
_{12}|u_{1}|^{2}|u_{2}|^{2}+\frac{1}{2}{\gamma }_{22}|u_{2}|^{4},  \label{h4}
\end{equation}%
where $\mathrm{c.c.}$ stands for the complex conjugate. The control
parameters of the system are $P$, $L$, $W$ and $\Omega $.

\section{Results and discussion}

Stationary solutions of Eqs.~(\ref{eq1}) and (\ref{eq2}) with propagation
constant $\beta$ were looked for as

\begin{equation}
u_{p}(x,y,z)=\phi _{p}(x,y)\exp (i\beta _{p}z),\quad p=1,2,  \label{beta}
\end{equation}%
where $\phi _{1,2}\left( x,y\right) $ are steady-state profiles, and real FF
and SH propagation constants are $\beta _{1}$ and $2\beta _{1},$
respectively. The stationary modes were found by means of the imaginary-time
propagation (ITP) method \cite{Weizhu,sodpre2006,fylpra2008} applied to Eqs.
(\ref{eq1}) and (\ref{eq2}).

To obtain stationary solutions by means of ITP, the inputs (at $z=0$) were
chosen as a vortex-type ansatz:

\begin{equation}
\phi _{1}(x,y;z=0)=r^{|S|}\exp (-r^{2})\exp (iS\theta ),
\end{equation}%
\begin{equation}
\phi _{2}(x,y;z=0)=r^{2|S|}\exp (-r^{2})\exp (2iS\theta ),
\end{equation}%
where $r$ and $\theta $ are the polar coordinates in the $\left( x,y\right) $
plane, and integer $S$ is the topological charge.

\begin{figure}[tbp]
	{\includegraphics[trim=115 15 90 15,clip,width=4.2in]{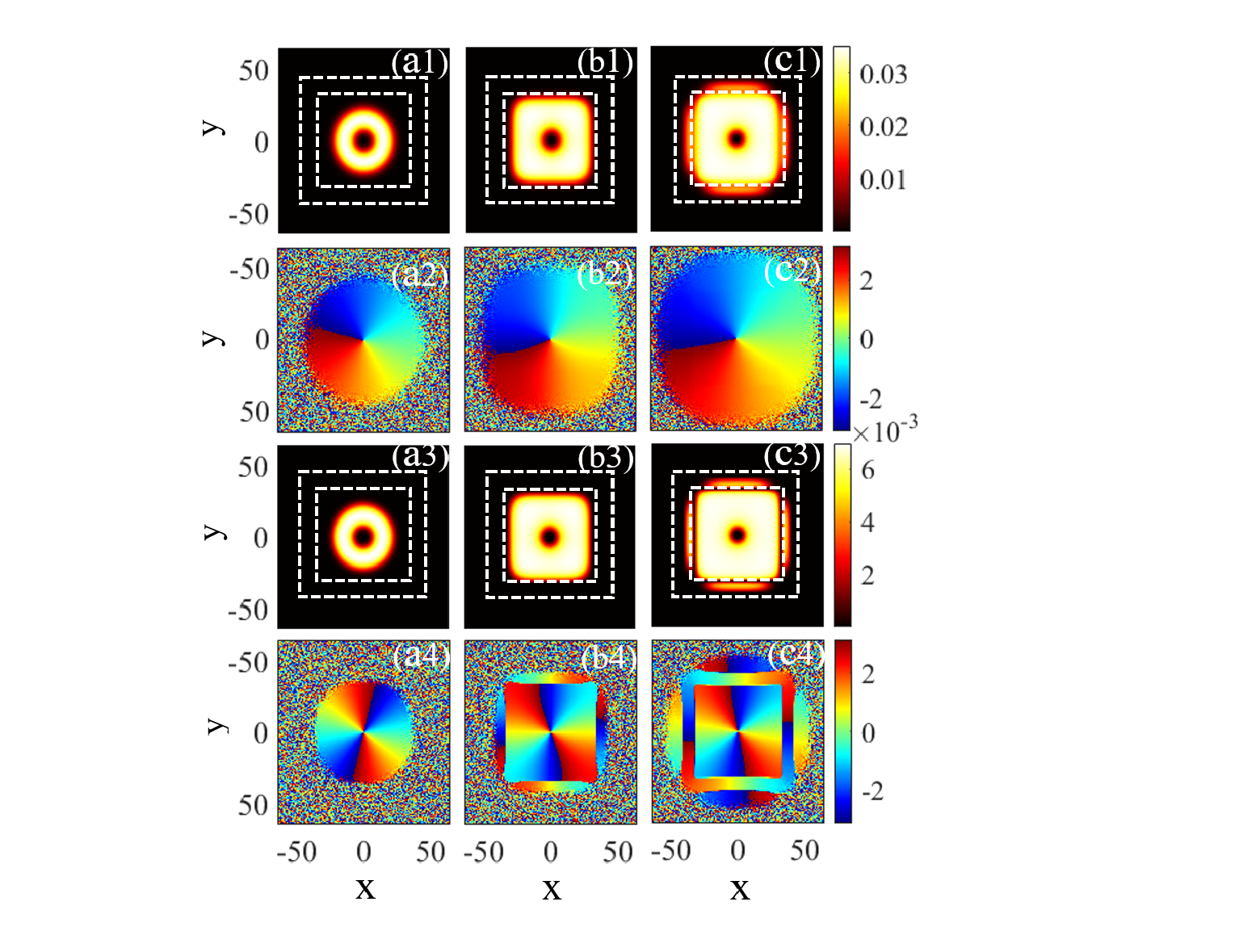}}
	\caption{Intensity and phase distributions of VD solitons. The white dashed
		squares indicate the boundaries of the modulation domains shown in Fig.~%
		\protect\ref{f1}(a). The phase discontinuities across these boundaries
		originate from the $\protect\pi $ phase difference (opposiite signs) of the
		nonlinearity coefficient. Panels (a1-a4) display the FF and SH components of
		the soliton with power $P=40$. Panels (b1-b4) and (c1-c4) correspond to $%
		P=120$ and the oversaturated VD mode with $P=200$, respectively. The
		parameters are $(L,W,\Omega )=(32,8,0)$. In direct simulations, all the
		modes are stable, at least, up to $z=10000$.}
	\label{f2}
\end{figure}

As expected, numerical results indicate that, with the increase of the total
power (\ref{power}), the transverse square-patterned modulation gradually
transforms rotationally symmetric VDs into square-shaped ones, which keep
the fourfold rotational symmetry ($C_{4}$) imposed by the modulation
structure. Upon further increase of $P$, the square-shaped VDs enter an
oversaturated regime and progressively exhibit outward expansion, Figure~\ref%
{f2} presents the intensity and phase distributions of the FF and SH
components of the VD solitons under the frequency-resonance condition, i.e.,
zero detuning. The white dashed square frames in Fig.~2 denote the
boundaries of the modulation domains introduced in Fig.~1(a). In these
domains, the nonlinearity coefficient $d(x,y)$ undergoes a sign reversal,
corresponding to a $\pi $ phase difference between adjacent regions. For the
relatively low total power in Figs.~\ref{f2}(a1) and~\ref{f2}(a3), the
competition between the $\chi ^{(2)}$ self-focusing and $\chi ^{(3)}$
defocusing nonlinearities gives rise to conventional annular vortex
solitons. In this regime, the phase distributions of both the FF and SH
components exhibit the well-defined phase circulations in Figs.~\ref{f2}(a2)
and~\ref{f2}(a4), indicating that the topological structure of the vortex
state remains intact.

As the total power is further increased in Figs.~\ref{f2}(b1) and~\ref{f2}%
(b3), the optical field gets strongly affected by the square-patterned
modulation, and the cubic defocusing nonlinearity gradually becomes the
dominant factor. Consequently, the original approximate rotational symmetry
of the annular VD, which was observed in Figs. ~\ref{f2}(a1) and~\ref{f2}%
(a3) is visibly broken, with both the FF and SH components evolving into
apparent square-shaped VDs, which keep solely the discrete fourfold
rotational symmetry. In this case, the square contours are characterized by
sharp edges and nearly uniform intensity distributions inside the squares.
With the subsequent increase in the total power in Figs.~\ref{f2}(c1) and~%
\ref{f2}(c3), the excessively strong defocusing cubic nonlinearity drives
the system into an oversaturated state, causing the square-shaped VDs to
gradually expand beyond the boundaries indicated by the white dashed lines.
Nevertheless, in this case too, the propagation of the square-shaped VDs is
completely stable. The distinctive spatial structures of the square VDs may
be of potential interest for applications to structured light generation and
nonlinear beam control \cite{Forbes,Nexus}

For the phase distributions of the VD solitons deployed in Fig.~\ref{f2},
the FF components exhibit phase circulation corresponding to the topological
charge $S=1$. This phase structure is maintained even when in
the oversaturated regime, indicating that the topological charge of the FF
component is protected in \ the course of the rotational-symmetry breaking
and saturation evolution of the VD modes. In parallel, the SH components of
the VD solitons maintains the phase circulation naturally corresponding to $%
S=2$.

\begin{figure}[tbp]
	{\includegraphics[trim=20 150 20 150,clip,width=\linewidth]{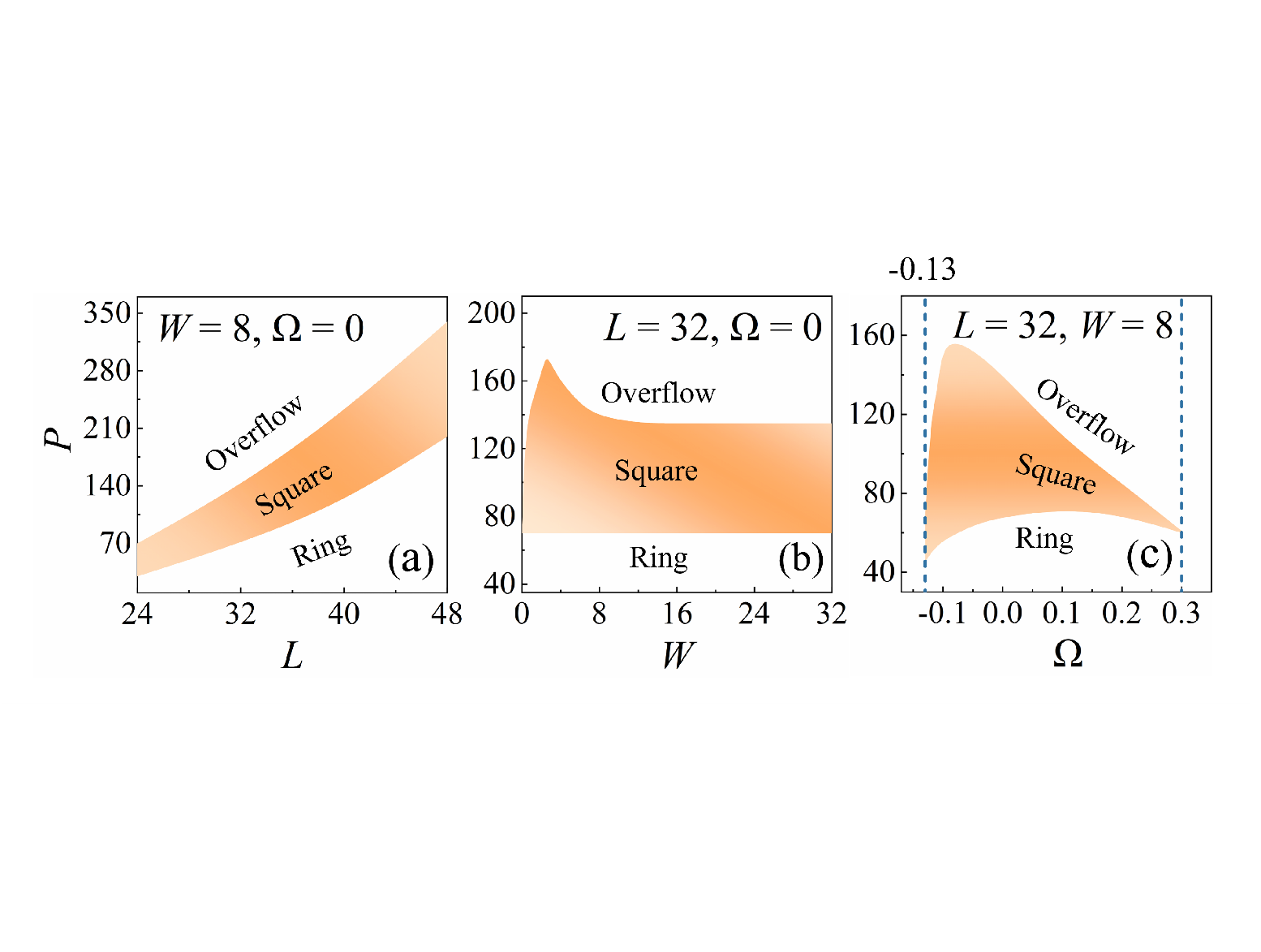}}
	\caption{In panels (a)-(c), the orange regions indicate the existence
		domains of stable square-shaped VDs in the $(P,L)$, $(P,W)$, and $(P,\Omega
		) $ parameter planes, respectively. The blank regions below the orange
		domains correspond to ring-shaped vortex states, whereas the blank regions
		above the domains represent the expanding overflow states. In panel (c), the
		regions outside of the dashed vertical lines do not support vortex
		solutions. }
	\label{f3}
\end{figure}

Next, we performed a systematic analysis of the existence domains of the
square-shaped VD solitons in the parameter space. In Fig.~\ref{f3}, the
existence regions of the stable VD solutions are plotted in the $(P,L)$, $%
(P,W)$, and $(P,\Omega )$ parameter planes. As seen in Fig.~\ref{f3}(a),
when $L$ is small and close to the lower bound, the parameter region capable
of supporting stable VD solutions in the $\left( P,L\right) $ plane is
essentially restricted. As $L$ increases, the corresponding transverse
confinement area provided by the square frame becomes larger, which
facilitates the stable formation of the VD states, leading to a
substantially expanded existence region. Next, by keeping the frame side
length $L$ fixed, we analyze the existence domain of the VD solutions in the
$(P,W)$ plane. As shown in Fig.~\ref{f3}(b), the existence region initially
expands with the increase of the frame thickness $W$, attains a maximum, and
then gradually decreases and eventually saturates. This behavior can be
attributed to the fact that a thin frame provides insufficient spatial
confinement for the optical field, whereas an excessively thick frame does
not enhance the confinement strength further, resulting in a nonmonotonous
dependence of the existence region on $W$. Finally, for fixed values of the
frame's side length $L$ and thickness $W$, we explore the existence domain
of the VD solutions in the $(P,\Omega )$ plane, as shown in Fig.~\ref{f3}%
(c). As $\Omega \to 0$, i.e., approaching the resonant condition, the nonlinear coupling efficiency is enhanced and the droplet existence region expands. In contrast, with increasing $|\Omega|$ (deviation from $\Omega = 0$), the coupling efficiency decreases and the existence region shrinks, resulting in a nonmonotonic dependence with a maximum near $\Omega = 0$.

To clarify the physical nature of the solutions outside the orange existence
domains in Fig. \ref{f3}, we performed a detailed analysis of the VD states
corresponding to the blank regions in the three parameter-space diagrams. As
shown in Fig.~\ref{f3}(a), in the blank regions above the orange domain, as
well as in all blank regions satisfying $L<24$, the solutions exhibit the
oversaturated regime, where, as mentioned above, the optical field extends
beyond the effective confinement imposed by the square frame. In contrast,
the blank regions located below the orange domain correspond to the
situation with a relatively low soliton's power, which supports not
square-shaped modes, but nearly axisymmetric stable annular ones, see Fig. %
\ref{f2}(a1-a4). A similar behavior is observed in the $(P,W)$ plane in Fig.~%
\ref{f3}(b). Specifically, the blank region above the orange domain is again
associated with oversaturated states, whereas the domain below the orange
domain corresponds to annular vortex solitons. In the $(P,\Omega )$ plane
shown in Fig.~\ref{f3}(c), square VDs exist only within a finite detuning
interval, namely $-0.13<\Omega <+0.3$, as denoted by the vertical dashed
lines. The blank regions outside the orange domain between the dashed
vertical lines, are populated, once again, by the oversaturated states in
the upper region and annular vortex solitons in the lower one, while no
vortex states are found outside of the stripe bounded by the vertical lines.

\begin{figure}[t]
\centering
\includegraphics[trim=55 120 20 110,clip,width=3.7in]{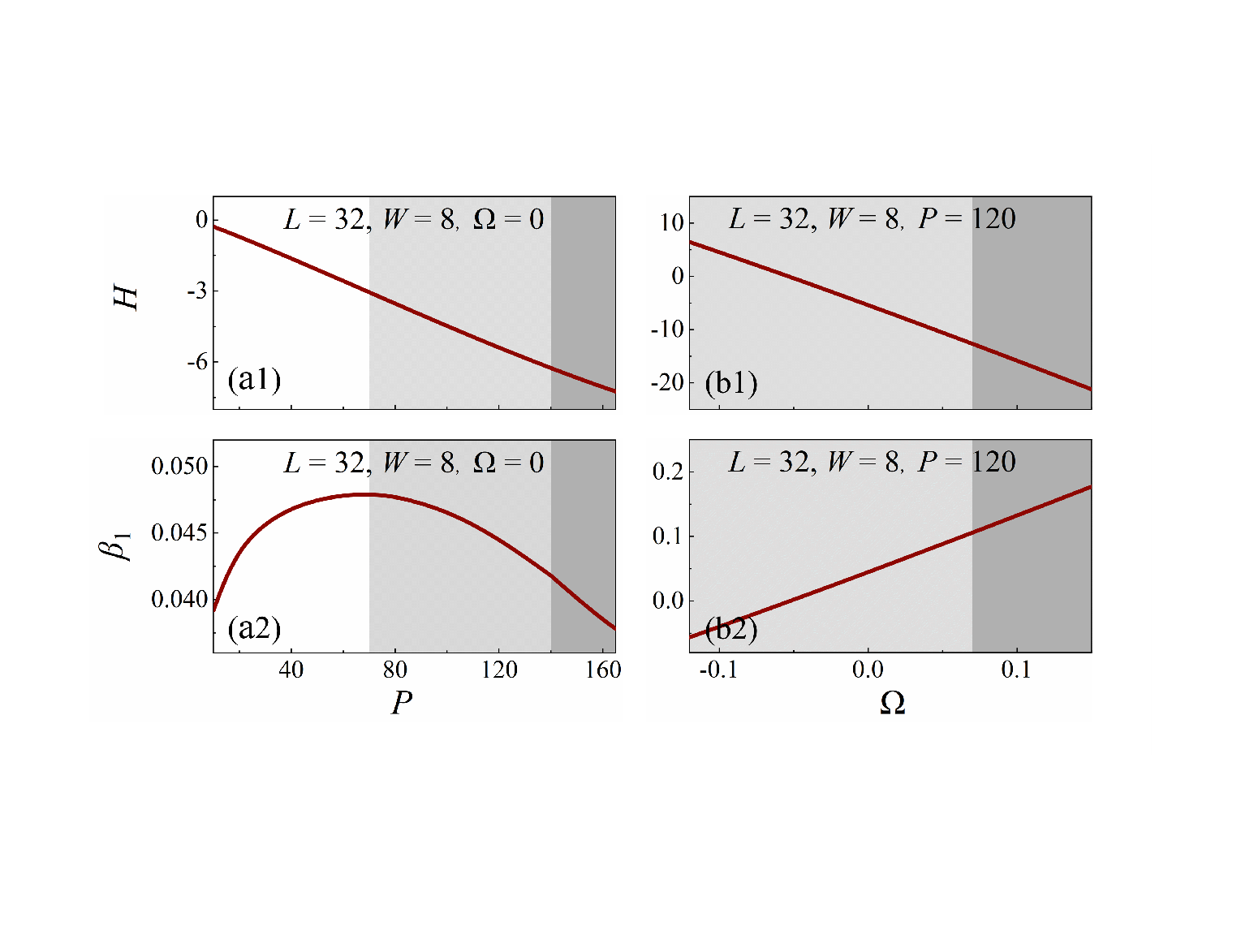}
\caption{The dependence of Hamiltonian $H$ and propagation constant $\protect%
\beta _{1}$ of the FF component on total power $P$ and detuning parameter $%
\Omega $. Panels (a1) and (a2) show $H(P)$ and $\protect\beta _{1}(P)$ at
fixed $\Omega =0$, respectively, while panels (b1) and (b2) display $%
H(\Omega )$ and $\protect\beta _{1}(\Omega )$ at fixed $P=120$. The system's
parameters are $L=32$ and $W=8$. The white, light-gray, and dark-gray
background regions are populated by the annular, square-shaped, and
overflow-type VD modes, respectively.}
\label{f4}
\end{figure}

Next, we analyze the dependence of Hamiltonian $H$ and propagation constant $%
\beta _{1}$ of the FF component of the square VDs on the total power $P$ and
detuning parameter $\Omega $. At $\Omega = 0$, with the geometric parameters of the square-patterned modulation
fixed as $L=32$ and $W=8$, the dependences are plotted in Figs.~\ref{f4}(a1)
and (a2). The white background region corresponds to $P<70$, where the
vortex soliton gradually evolves from the annular shape into a square VD, as
the total power increases. In this regime, the dependence $\beta _{1}(P)$
satisfies the Vakhitov--Kolokolov (VK) stability criterion, $d\beta
_{1}/dP>0 $ \cite{nqe1973}. A clearly developed square-shaped VD emerges at $%
P\simeq 70 $, which is consistent with the minimum power threshold for the
existence of the same modes, which are identified in Figs.~\ref{f3}(a) and
(b). At $70<P<140$, the light-gray background region indicates that the
square-shaped VDs persist without entering the above-mentioned overflow
regime, their square profiles getting progressively sharper with the
increase of $P$. In this power interval, the \emph{self-defocusing} cubic
nonlinearity is the dominant term in Eqs. (\ref{eq1}) and (\ref{eq2}). As a
result, the system exhibits an \emph{anti-VK-type} behavior, with $d\beta
_{1}/dP<0$. This conclusion is consistent with the known fact that the
anti-VK condition, $d\beta _{1}/dP<0$, is a necessary stability criterion
for localized modes supported by a self-repulsive nonlinearity (e,g,, gap
solitons in other models) \cite{HS}. Further, when the parameters are fixed
as $(L,W,P)=(32,8,120)$, the dependences of the Hamiltonian and propagation
constant on the detuning parameter are shown in Figs.~\ref{f4}(b1) and \ref%
{f4}(b2). We find that the admissible detuning range for the existence of
the stable square-shaped VDs is $-0.12<\Omega <+0.07$. This interval agrees
well with the existence domain which can be identified in the $(P,\Omega )$
plane shown in Fig.~\ref{f3}(c), thereby further confirming the consistency
of the parameter-space analysis. It is worth emphasizing that, in the
overflow regime, indicated by the dark-gray background region in Fig. \ref%
{f4}, the Hamiltonian attains its lowest values; however, this fact does not
imply enhanced stability. Instead, it reflects the loss of the spatial
localization, whereby the system reduces its energy by spreading the field
distribution beyond the square-patterned modulation area.

\begin{figure}[t]
\centering
\includegraphics[trim=105 125 100 75,clip,width=3.7in]{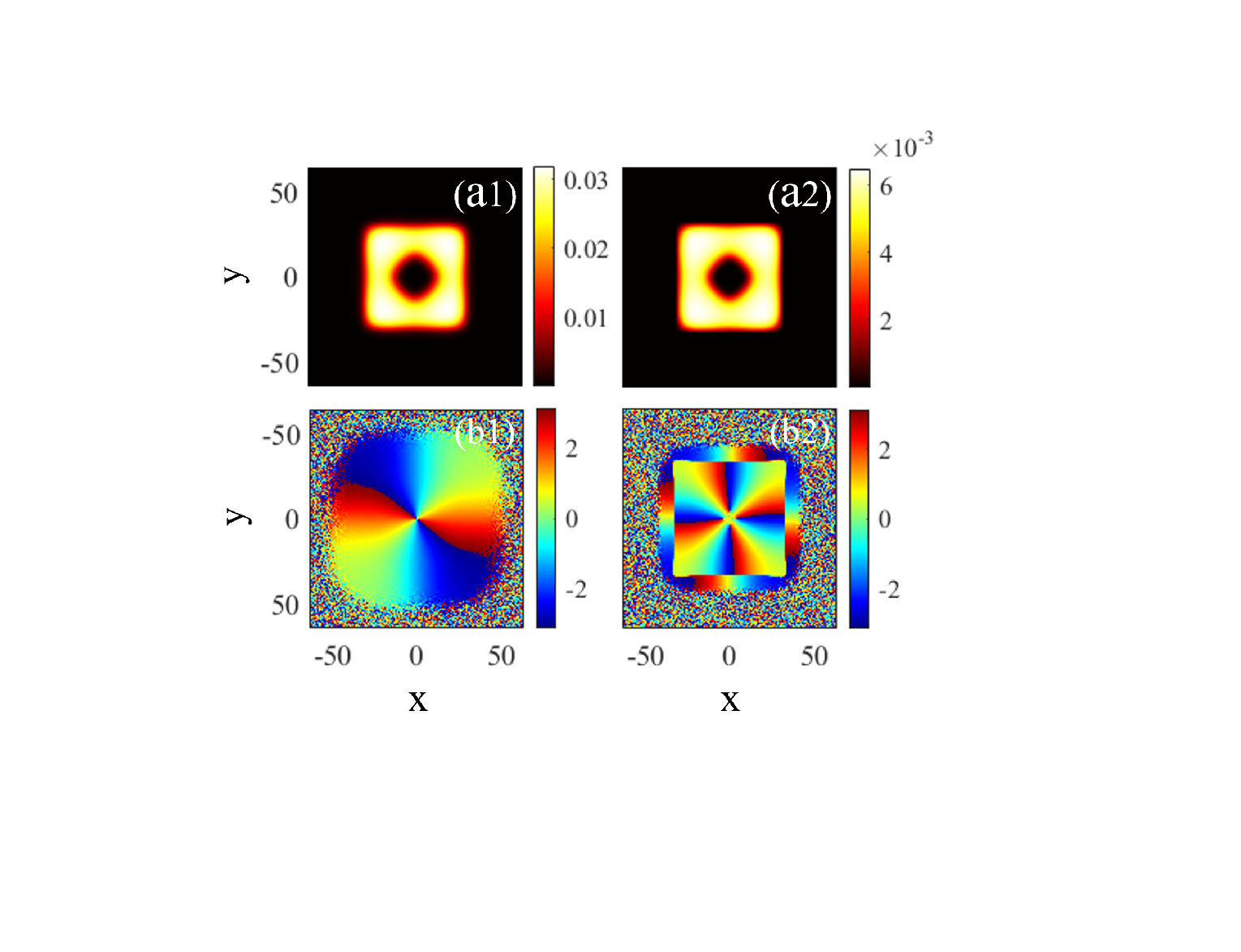}
\caption{Examples of stable square-shaped higher-order VDs with topological
charges $S=2$ (panels (a1), (b1)) and $S=4$ (panels (a2), (b2)). Panels
(a1), (a2) and (b1), (b2) display the intensity and phase distributions of
the FF and SH components, respectively. The VDs with $S=2$ remain stable in
simulations for the propagation distance, at least, $z=10000$. The other
parameters are $(P,L,W,\Omega )=(120,32,8,0)$. }
\label{f5}
\end{figure}

\begin{figure}[t]
\centering
\includegraphics[trim=105 105 100 90,clip,width=3.6in]{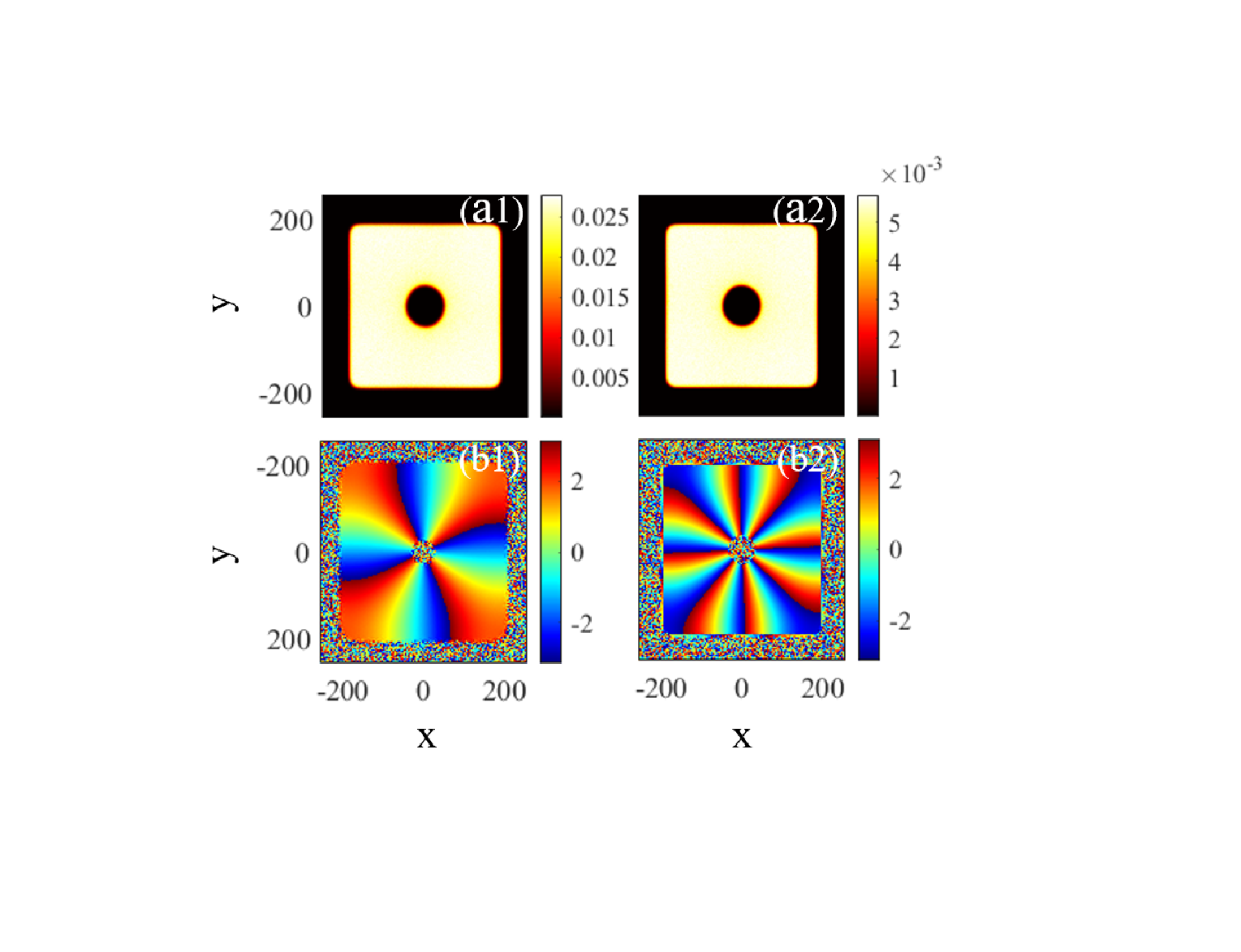}
\caption{Examples of stable square-shaped higher-order VDs with topological
charges $S=4$ (panels (a1), (b1)) and $S=8$ (panels (a2), (b2)). Panels
(a1), (a2) and (b1), (b2) display the intensity and phase distributions of
the FF and SH components, respectively. The vortex droplets with $S=4$
remain stable in simulations for the propagation distance, at least, $z=10000
$. The other parameters are $(P,L,W,\Omega )=(5000,192,24,0)$. }
\label{f6}
\end{figure}

The generality of the shaping mechanism is further demonstrated by
considering vortex solitons with higher values of the topological charge $S$%
, supported by the system. Figures~\ref{f5} and \ref{f6} display
representative examples of square-shaped higher-order VD modes with $S=2$
and $S=4$, and $S=4$ and $S=8$, respectively. In both figures, panels (a1)
and (a2) show the intensity distributions of the square-shaped VDs for the
fundamental-frequency (FF) and second-harmonic (SH) components with the
corresponding topological charges, while panels (b1) and (b2) present the
associated phase patterns. Under the same modulation parameters as those
used in Fig.~\ref{f2}, the stable existence region for these higher-order
vortex solitons with $S=2$ and $S=4$ is found to be $100<P<140$. For even
higher-order vortex states, e.g., $S=4$ and $S=8$, respectively, stable
square-shaped configurations can only be produced when the size of the
modulation square is, at least ,three times larger than that used in Fig.~%
\ref{f2}, indicating that broader spatial confinement is required to
accommodate the increased phase winding. On the other hand, no stable square-shaped
vortex droplet solutions are found for $S=3$. This fact can be attributed to
the mismatch between the intrinsic fourfold rotational symmetry of the
underlying modulation structure and the topological charge $S=3$, which
prevents the formation of an appropriate phase structure. More generally, a given discrete rotational symmetry tends
to support vortex states whose topological charge is compatible with that symmetry,
while hindering the formation of states with mismatched topology. All the VD
solutions shown here are found to be stable. These results demonstrate
that the transverse square modulation enables the shaping and stabilization
not only of fundamental vortex solitons with $S=1$, but also of higher-order
ones with $S=2$, $4$, and $8$. However, for shaping the higher-order
vortices, a larger modulation square-shaped area is required.

\section{Estimates of experimental parameters}

To assess the experimental feasibility of the proposed species of vortex
solitons, we performed a systematic quantitative analysis based on material
parameters of lithium niobate (LiNbO$_{3}$). Accordingly, the second-order
nonlinear coefficient and cubic nonlinear susceptibility in Eq.~(\ref{d1})
are taken as $d_{0}=d_{22}=2.1~\mathrm{pm/V}$ \cite{vhn64} and $\chi
^{(3)}=36.6\times 10^{-22}~\mathrm{m^{2}/V^{2}}$ \cite{ispinjose}. The
wavelengths of the FF and SH components are chosen as $\lambda _{1}=1064~%
\mathrm{nm}$ and $\lambda _{2}=532~\mathrm{nm}$, respectively. The
refractive indices of the two components are approximately equal, $%
n_{1}\approx n_{2}\approx 2.2$. Under these conditions, the conversion
relations between the normalized variables and the corresponding physical
quantities are summarized in Table~\ref{tab:relation}. According to Eq.~(\ref%
{zd}), the characteristic propagation distance, $z=10\,000$, corresponds to
a physical length $\approx 2.8~\mathrm{m}$, which is much larger than the
underlying diffraction length, making the stability prediction reliable \cite%
{kprl2017,jprl2003}. Based on the parameters listed in Table~\ref%
{tab:relation}, the peak intensities of the FF and SH components of the
square VD soliton shown in Fig.~\ref{f2} are estimated to be $\approx 4~%
\mathrm{GW/cm^{2}}$ and $0.43~\mathrm{GW/cm^{2}}$, respectively. Note that
the spatial scales of the structures listed in Table~\ref{tab:relation} fall
well within the fabrication capabilities of currently available QPM
techniques. These results not only support the experimental feasibility of
the proposed scheme, but also provide clues for the optimization of
parameters in potential applications.

\begin{table}[h]
\caption{Relations between the scaled variables and the corresponding
physical units of the coordinates, total power, and intensity}
\label{tab:relation}\centering
\setlength{\arrayrulewidth}{0.8pt} 
\begin{tabular}{cc}
\hline\hline
$x = 1\,\&\, y = 1$ & 4.64 $\mu$m \\
$z = 1$ & 280 $\mu$m \\
$P = 1$ & 31 kW \\
$|u_1|^2 = 0.01\,\&\,|u_2|^2 = 0.01$ & \hspace{2em}1.44 \,\&\, 0.72 GW/cm$^2$
\\ \hline\hline
\end{tabular}%
\end{table}

\section{Conclusion}

We have systematically investigated the controllable mechanism for the
recasting of usual annular vortex solitons into square-shaped VDs (vector
droplets) in QPM (quasi-phase-matched) photonic crystals by introducing the
square pattern of the transverse modulation. Systematic simulations
demonstrate that, with the increase of the total power, the interplay
between the quadratic self-focusing and cubic defocusing in the photonic
crystal is gradually biased towards the domination of the cubic term. As a
consequence, the continuous rotational symmetry inherent to conventional
(annular) vortex solitons is broken, leading to the formation of
square-shaped VD modes, featuring the discrete fourfold rotational symmetry (%
$C_{4}$). These square-shaped VDs preserve the well-defined topological
phase circulation, simultaneously exhibiting robust self-trapping dynamics
in the course of the propagation.\ Systematically scanning the parameter
space, we have identified the existence domains of the square-shaped VD
solutions in terms of the total power, side length and thickness of the
square modulation frame, and the detuning parameter. Distinct regions
corresponding to the low-power vortex rings, square-shaped VDs, and
expanding oversaturated states are identified in the parameter space. The
results demonstrate the sturdiness of the vortex solitons, and reveal that
optimal matching of the parameters of the square-pattern modulation and
nonlinearity strength significantly expands the existence domain of the
stable square-shaped VDs, whereas insufficient or excessive confinement is
unfavorable for their formation. In the course of the formation of the
square-shaped VDs, the dependence of the propagation constant on the total
power switches from satisfying the VK (Vakhitov-Kolokolov) criterion to the
anti-VK one, in the regimes dominated by the quadratic and defocusing cubic
nonlinear terms, respectively. Estimates based on material parameters of
lithium niobate suggest that the square-shaped VD modes produced in this
work are feasible under currently available technological conditions.

In summary, the present results show that the appropriate spatial modulation
of the QPM photonic crystal enables controllable reshaping of vortex
solitons into square VDs, offering the strategy for the design of sturdy
optical modes with tailored symmetry and topology. At a more fundamental level, our findings reveal the unexpected universality: the vortex droplets retain their topological charge and robust propagation when their spatial symmetry is reduced from the continuous rotational invariance to the discrete-square form. These findings demonstrate that the essential assets of the vortex modes are not contingent upon the circular symmetry. More generally speaking, the use of other shapes of the transverse modulation, such as hexagonal or triangular, may serve as a platform for testing the generality of this principle in the context of other discrete symmetries, such as $C_{6}$ or $C_{3}$. These studies offer new options for the design of structured light patterns, while further probing limits of the topological robustness under broken symmetries.

\section*{Acknowledgments}

This work was supported by NNSFC (China) through Grants No. 12274077 and
12475014, the Natural Science Foundation of Guangdong province through Grant
No. 2025A1515011128, No. 2024A1515030131, No. 2023A1515110198, and No.
2023A1515010770, the Research Fund of Guangdong-Hong Kong-Macao Joint
Laboratory for Intelligent Micro-Nano Optoelectronic Technology through
grant No.2020B1212030010.

\end{document}